\documentclass[floatfix,citeautoscript,preprint,11pt,superscriptaddress]{revtex4-1}

\usepackage{graphicx}
\usepackage{subfigure}
\usepackage{setspace}[1.5] 
\usepackage{amsmath, amsthm, amssymb}
\usepackage{xfrac}
\usepackage{dcolumn}
\usepackage{gensymb}
\usepackage{threeparttable}
\usepackage[version=3]{mhchem} 
\usepackage{verbatim}

\usepackage{chemfig}
\usepackage{chemformula}    

\usepackage{listings}

\usepackage{color}

\begin{document}

\date{\today} 

\title{  
Long range ionic and short range hydration effects govern strongly anisotropic clay nanoparticle interactions 
}

\author{Andrea Zen}\thanks{These two authors contributed equally}
\affiliation{Dipartimento di Fisica Ettore Pancini, Universit\`{a} di Napoli Federico II, Monte S. Angelo, I-80126 Napoli, Italy}
\affiliation{Department of Earth Sciences, University College London,  Gower Street, London WC1E 6BT, United Kingdom}
\affiliation{Thomas Young Centre and London Centre for Nanotechnology, 17--19 Gordon Street, London, WC1H 0AH, United Kingdom}
\email{andrea.zen@unina.it}

\author{Tai Bui}\thanks{These two authors contributed equally}
\affiliation{Thomas Young Centre and London Centre for Nanotechnology, 17--19 Gordon Street, London, WC1H 0AH, United Kingdom}
\affiliation{Department of Physics and Astronomy, University College London,  Gower Street, London WC1E 6BT, United Kingdom}
\affiliation{BP Exploration Operating Co. Ltd, Chertsey Road, Sunbury-on-Thames TW16 7LN, United Kingdom}

\author{Tran Thi Bao Le}
\affiliation{Department of Chemical Engineering, University College London, WC1E 7JE, London, United Kingdom}

\author{Weparn J. Tay}
\affiliation{BP Exploration Operating Co. Ltd, Chertsey Road, Sunbury-on-Thames TW16 7LN, United Kingdom}

\author{Kuhan Chellappah}
\affiliation{BP Exploration Operating Co. Ltd, Chertsey Road, Sunbury-on-Thames TW16 7LN, United Kingdom}

\author{Ian R. Collins}
\affiliation{BP Exploration Operating Co. Ltd, Chertsey Road, Sunbury-on-Thames TW16 7LN, United Kingdom}

\author{Richard D. Rickman}
\affiliation{BP Exploration Operating Co. Ltd, Chertsey Road, Sunbury-on-Thames TW16 7LN, United Kingdom}

\author{Alberto Striolo}
\affiliation{Department of Chemical Engineering, University College London, WC1E 7JE, London, United Kingdom}
\affiliation{School of Chemical, Biological and Materials Engineering, University of Oklahoma, Norman, OK 73019, United States} 

\author{Angelos Michaelides}
\affiliation{Yusuf Hamied Department of Chemistry, University of Cambridge, Lensfield Road, Cambridge CB2 1EW, United Kingdom}
\affiliation{Thomas Young Centre and London Centre for Nanotechnology, 17--19 Gordon Street, London, WC1H 0AH, United Kingdom}
\affiliation{Department of Physics and Astronomy, University College London,  Gower Street, London WC1E 6BT, United Kingdom}
\email{am452@cam.ac.uk}

\begin{abstract}
The aggregation of clay particles in aqueous solution is a ubiquitous everyday process of broad environmental and technological importance. 
However, it is poorly understood at the all-important atomistic level since it depends on a complex and dynamic interplay of solvent-mediated electrostatic, hydrogen-bonding, and dispersion interactions.
With this in mind we have performed an extensive set of classical molecular dynamics simulations (included enhanced sampling simulations) on the interactions between model kaolinite nanoparticles in pure and salty water. 
Our simulations reveal highly anisotropic behaviour in which the interaction between the nanoparticles varies from attractive to repulsive depending on the relative orientation of the nanoparticles. 
Detailed analysis reveals that at large separation ($>$1.5 nm) this interaction is dominated by electrostatic effects whereas at smaller separations the nature of the water hydration structure becomes critical. 
This study highlights an incredible richness in how clay nanoparticles interact, which should be accounted for in e.g. coarse grained models of clay nanoparticle aggregation. 
\end{abstract}

\maketitle



\section{Introduction}

Clay particles are all around us: present in the Earth's crust, soil, the ocean floor, and the atmosphere as aerosols.
Clays have been extensively used since antiquity in pottery, as building and writing materials, and more. 
Clays are also critical to contemporary challenges: For example, as atmospheric ice nucleating agents they are relevant to climate change; the ability of clays to trap toxic (including nuclear) waste is important for environmental remediation; and as porous materials they filter and dictate the flow of water and other fluids through rocks. 

Understanding the aggregation of individual clay particles into larger agglomerates is critical to explaining and controlling the behaviour of clays in many of the above examples. 
However, at the atomic scale, clay particle agglomeration is poorly understood.
Partly this is down to the challenge of characterising and tracking the agglomeration of clay particles in solution but also because of the wealth of variables that are inevitably relevant to the agglomeration process. 
Parameters such as particle size, shape, chemical composition, temperature, pressure, and solvent effects will all play a role. 
A further challenge is the inherent chemical complexity of clay particles: as (alumino)-silicates clays interact through a complex interplay of electrostatic, hydrogen bonding, and van der Waals dispersion forces; with all these interactions mediated by the aqueous electrolyte solution separating individual particles.

The inherent (and interesting) complexity of clay particle aggregation has motivated a large body of work aimed at understanding clay particle association under well-defined conditions, as well as related questions about the structure and dynamics of the water-clay interface \cite{Siretanu2014, Kumar2016, Kumar2017, Gupta2010, Gupta2011, Liu2014, Dishon2009, Tombacz2006, Gan2006, Hu1995, Furukawa2009a,
Volkova2021, Shen2021MolecularWater, Ho2021Molecular-levelWater, Ebrahimi2012, Ebrahimi2014, Ebrahimi2016, Liu2012a, Liu2015, 
Vasconcelos2007, Li2015, Li2016, Zhang2016, Underwood2016, Papavasileiou2018, Sayer2020b,
Tazi2012a, Liu2013, Liu2014_pka_21clay,
Hu2008, Presti2016, Liao2019, Pouvreau2019a}. 
Being the simplest, one of the most abundant, and also one of the most technologically relevant clays, kaolinite (Al$_2$(OH)$_4$Si$_2$O$_5$), 
has emerged as a widely studied model system. 
Indeed kaolinite can now essentially be considered the ``fruit fly'' system of physical chemistry / chemical physics research on clay particles. 
However, the nature of kaolinite nanoparticle association under even simple and well-defined conditions is still not understood. 
This is what we aim to address in the current study through a systematic set of computational studies aiming specifically to understand the following questions: 
(i) Are the interactions between clay (nano)-particles in solution attractive or repulsive?; 
(ii) Does the nature of the interaction depend on the facets of the (nano)-particles that interact and if so to what extent?; 
(iii) What role does the aqueous solution play in the association process?; 
and (iv) What are the length scale(s) of the particle-particle interactions?

To answer these questions and to gain a better understanding of the factors that influence kaolinite particle agglomeration in aqueous solutions, we used molecular dynamics simulations to explore the association of clay nanoparticles. 
Since kaolinite nanoparticles most often occur as hexagonal platelets \cite{kao_particle_hex}, we modelled the association of 
hexagonal clay nano-platelets. 
In order to capture the role the (dynamic) solvent molecules play in the association process we computed potential of mean force (PMF) curves between two kaolinite particles with varying orientations in both pure and saline water environments. The contributions of interparticle interactions are disentangled into the van der Waals and Coulomb potentials. The effect of the hydration films on the PMF profiles is then investigated, with a particular emphasis on the structural and dynamical properties of water molecules confined between the two approaching surfaces. As the presence/absence of ions in solution affects differently the different contributions determining the overall interparticle interactions, we are able to elucidate the effects of salt on the evaluated PMFs. 

The remainder of the paper is organised as follows. In section \ref{sec:methods} we describe the approach used to model kaolinite particles, and the computational details of the performed simulations.
In section \ref{sec:results} we present and discuss the results of our simulations,  highlighting the orientational dependence of the interaction among nanoparticles, and showing how qualitative features can be rationalised in terms of hydration and electrostatic charge at the interface. 
Conclusions are finally given in section \ref{sec:conclusions}.
Additional material on the simulation techniques, the convergence tests performed, and some further analysis are reported in the supporting information (SI).

\section{Theory and Computational Details}\label{sec:methods}

\begin{figure}[ht]
\centering
\includegraphics[page=1, width=\textwidth, trim=0cm 0cm 0cm 0cm,clip]{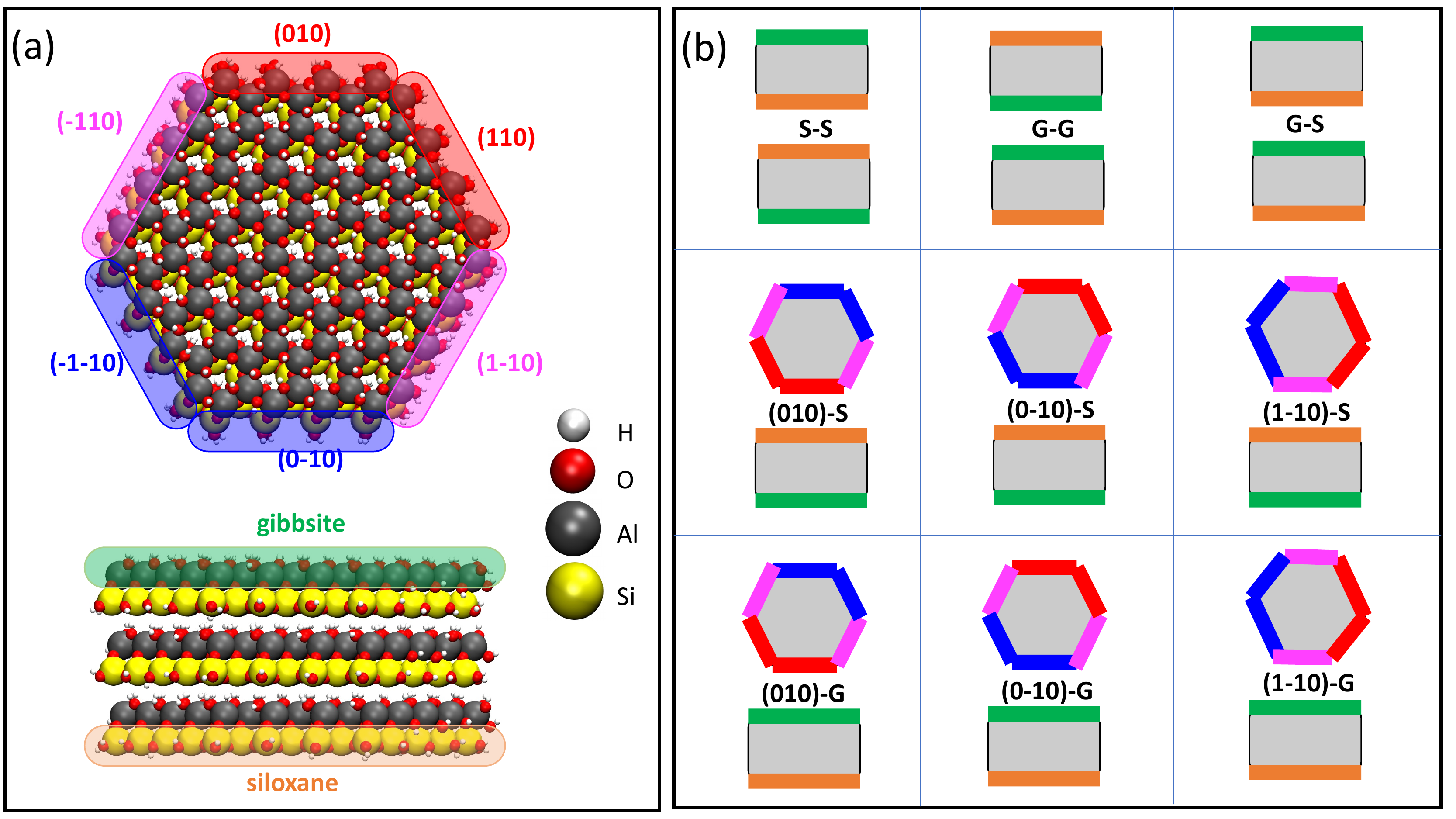}
\caption{\label{fig_MD}
Overview of the model kaolinite nanoparticles considered in this study. Panel (a) shows a representation of two kaolinite particles in the edge-to-face orientation. The six edges (top) are called after their Miller indices of the crystallographic structure \cite{Bish1993}, and the two inequivalent basal surfaces (bottom), (001) and (00-1), are called gibbsite (G) and siloxane (S) faces, respectively, according to the literature.
White, red, grey and yellow spheres represent the hydrogen, oxygen, aluminium  and silicon atomic species, respectively. We emphasise the 5 nonequivalent surfaces (the six edges are 2-by-2 equivalent) using different colours.  Panel (b) shows the nine representative orientations of two kaolinite particles considered in this work. 
}
\end{figure}

The MD simulations employing force field potentials have all been performed using the LAMMPS package \cite{Plimpton1995}. 
%
In each simulation cell there are two kaolinite nanoparticles and 10,000 to 11,000 water molecules. 
An image of the nanoparticles in edge to face configuration -- but without any water present -- is shown in Figure \ref{fig_MD}a. An image of the full simulation cell including the water molecules is shown in  Figure~S1 
of the Supporting Information (SI). %
%
The kaolinite nanoparticles are prepared as described in section~S1 
of the SI. To simulate the saline solution, sodium chloride ions corresponding to the salinity of 1.2 M were added to the system. 
In line with ref. \cite{Ho2021Molecular-levelWater}, the simulation cell is roughly $60\times 60\times 100$~\AA$^3$. 
%
A simulation box of this size is sufficiently large to avoid finite size effects, as evidenced by: (i) the density of water, see Figure \ref{fig3}c and Figure~S10g-i 
of the SI; and (ii) the diffusion coefficient of water, see Figure~S13 
of the SI, in the region between the two kaolinite particles, both of which reach the bulk value. 
%
%
In the face-to-face orientation both the nanoparticles have the basal faces parallel to the $xy$-plane, while in the face-to-edge orientation one nanoparticle (the lower in our set-ups) has the basal faces parallel to the $xy$-plane, and the other nanoparticle has the basal faces orthogonal to the $xy$-plane and it is placed above the first nanoparticle. 

\subsection{Force fields}\label{sec:force_fields}
The simulations employed the ClayFF potential \cite{Cygan2004d, Pouvreau2017, Pouvreau2019a} for kaolinite and the rigid SPC/E model \cite{Berendsen1987c} for water. The sodium, Na$^+$, and chloride ions, Cl$^-$, were modeled as a single charged Lennard-Jones (LJ) sphere with parameters taken from the study of Smith and Dang without polarizability \cite{Smith1994a}.
Interactions between unlike atom types were calculated using the standard Lorentz-Berthelot mixing rules. 
The SHAKE algorithm \cite{Ryckaert1977} is used to constrain the rigid water molecule and the OH distance in the hydroxyl groups of kaolinite.
Additional constraints on the kaolinite atoms have been added to enhance the stability of the nanoparticles, following previous work \cite{Ho2017, Ho2018}.
Real-space interactions were truncated at 10~\AA{} with corrections to the energy applied, and a particle-particle-particle-mesh (PPPM) solver was used to account for long-range electrostatics \cite{Hockney_1988}.
The integration step in the MD simulations is 1~fs. 
We tested the above setup on the adsorption of a single water molecule, of the Na$^+$ cation and Cl$^-$ anion on the siloxane and gibbsite faces, and we registered reasonable agreement with benchmark {\em ab-initio} evaluations (see 
section~S2 
and Table~S1 
in the SI), in agreement with other studies from literature \cite{Zen2016_kao, Pouvreau2019a, Presti2016}.

\subsection{Simulation protocol}\label{sec:sim_protocol}

To simulate the system at the desired temperature and pressure, an equilibrium simulation at a pressure of 200 bar and temperature of 350~K
was initially conducted for 1~ns in the NPT ensemble (constant number (N) of atoms/particles, pressure (P), and temperature (T)).
Moreover, the system associated to each specific orientation has been equilibrated in the NVT ensemble (constant number of atoms/particles, volume (V), and temperature) at 350 K for over 1 ns by applying separate Nos\'{e}-Hoover chain thermostats \cite{Nose1984} to water and nanoparticles.

The umbrella sampling (US) technique\cite{Torrie1974MonteFluid, Torrie1977NonphysicalSampling} was employed to evaluate the PMF among two nanoparticles using the collective variables module (COLVARS) \cite{Fiorin2013} implemented in LAMMPS \cite{Plimpton1995}.
%
The computational details for the US simulations are reported in the SI. In these simulations, a harmonic potential with a force constant of 10 kcal/(mol.m$^2$) or 50 kcal/(mol.m$^2$) is used to keep the upper particle at a specific distance from the bottom one to ensure good sampling overlap between adjacent sampling windows. In both the face-to-face and face-to-edge orientations, the lower nanoparticle is not allowed to diffuse nor to rotate, with its heavy atoms (Al and Si) tethered to their initial position (using the {\em fix spring/self} command in LAMMPS).
The upper nanoparticle is allowed to translate but not to rotate (heavy atoms are constrained using the {\em fix rigid} command). The only exception is for the cases of the edge-to-face orientations, where a short simulation (around 100 ps, prior to imposing the rotational constraints) was performed to allow the upper particle to freely rotate and translate so that the edge of the top particle can preferentially interact with the basal surface of the bottom one. Consequently, the angle formed by the two particles are not precisely 90 degrees afterwards (the deviation is less than 10 degrees). 
Note that the PMF is a function of the distance among the nanoparticles.
There are several different ways to quantify this distance (see for instance  Figure~S1 
in the SI).
The US constraint uses the distance among the centres of mass of the heavy atoms (i.e., Al and Si) of the two nanoparticles, as it is easily computed in runtime during an MD simulation. 
In the results section we report the results as a function of the minimum distance $\Delta Z$ among the heavy atoms of the two nanoparticles, which provides a direct indication of the thickness of the interface and facilitates the comparison among PMF curves relative to different orientations.
%
%
We note that the procedure described above was followed for each orientation of the nanoparticles shown in Figure~\ref{fig_MD}, panel b, and discussed in the results.

As noted above, the simulations reported in this work have been performed at 350~K, representing a typical temperature for inner earth oil reservoirs. 
We also performed additional simulations at 450~K on the gibbsite-siloxane orientation 
(see section S4 
and figure S5 
in the SI).
The effect of temperature on the PMF profile yields a minimal impact.
This observation is in agreement with the observations of \citet{Ho2021Molecular-levelWater}.
%
Finally, it is important that the US MD simulations are long enough to sample properly the phase space.
We performed some preliminary calculations, which indicated that a reasonable level of convergence in the PMF is already achieved when each US MD simulation is 3 ns long or more, 
see Figure~S4 
of the SI. 
In our production calculations, we used at least 5 ns long US simulations.

\section{Results and discussion}\label{sec:results}

In the following sections we first show and discuss the PMFs between kaolinite nanoparticles at different orientations, section 3.A, 
followed by an analysis of the energetic contributions of the free energy and how it determines the attractive or repulsive nature of the interaction between nanoparticles, section 3.B. 
In section 3.C 
we analyse the structure of water and ions at the interface and elucidate how they determine the qualitative features of the PMF profile for narrow separation distances.

\subsection{Orientation dependence of potential of mean force profiles}\label{sec:PMFs}

A hexagonal kaolinite platelet has eight interfaces with water: two inequivalent basal faces and six edges. 
A careful analysis of the six edges shows that some of them are indeed very similar, and we can identify only 3 inequivalent edges \cite{Pouvreau2019a, White1988}.
In particular, the (0 1 0) edge is roughly equivalent to the (1 1 0), the (0 -1 0) edge is roughly equivalent to the (-1 -1 0) surface, and the (1 -1 0) edge is roughly equivalent to the (-1 1 0) surface, see Figure \ref{fig_MD}.
Thus, we limit our investigation to only one representative edge for each equivalent class.
This leads to three representative edges: (0 1 0), (1 -1 0), and (0 -1 0).
Therefore, in aqueous solutions, we evaluated nine distinct PMFs for particle-particle interactions, three of which are face-to-face oriented and six of which are edge-to-edge oriented. 
%
%
In real systems, we expect the presence of a variety of ions in water, either directly desorbed from rock surfaces or in the form of mineral salts. 
In addition, given that the kaolinite basal surface is polar \cite{Hu2010}, ions in solution are likely to adsorb at the interface to compensate the dipoles present within the nanoparticle \cite{Sayer2020b}.  
%
To account for these issues we 
performed simulations in both pure water and in the more realistic situation of saline water, with a 1.2M sodium chloride (NaCl) concentration.

\begin{figure}[ht!]
\centering
\includegraphics[width=\textwidth, trim=0cm 0cm 0cm 0cm,clip]{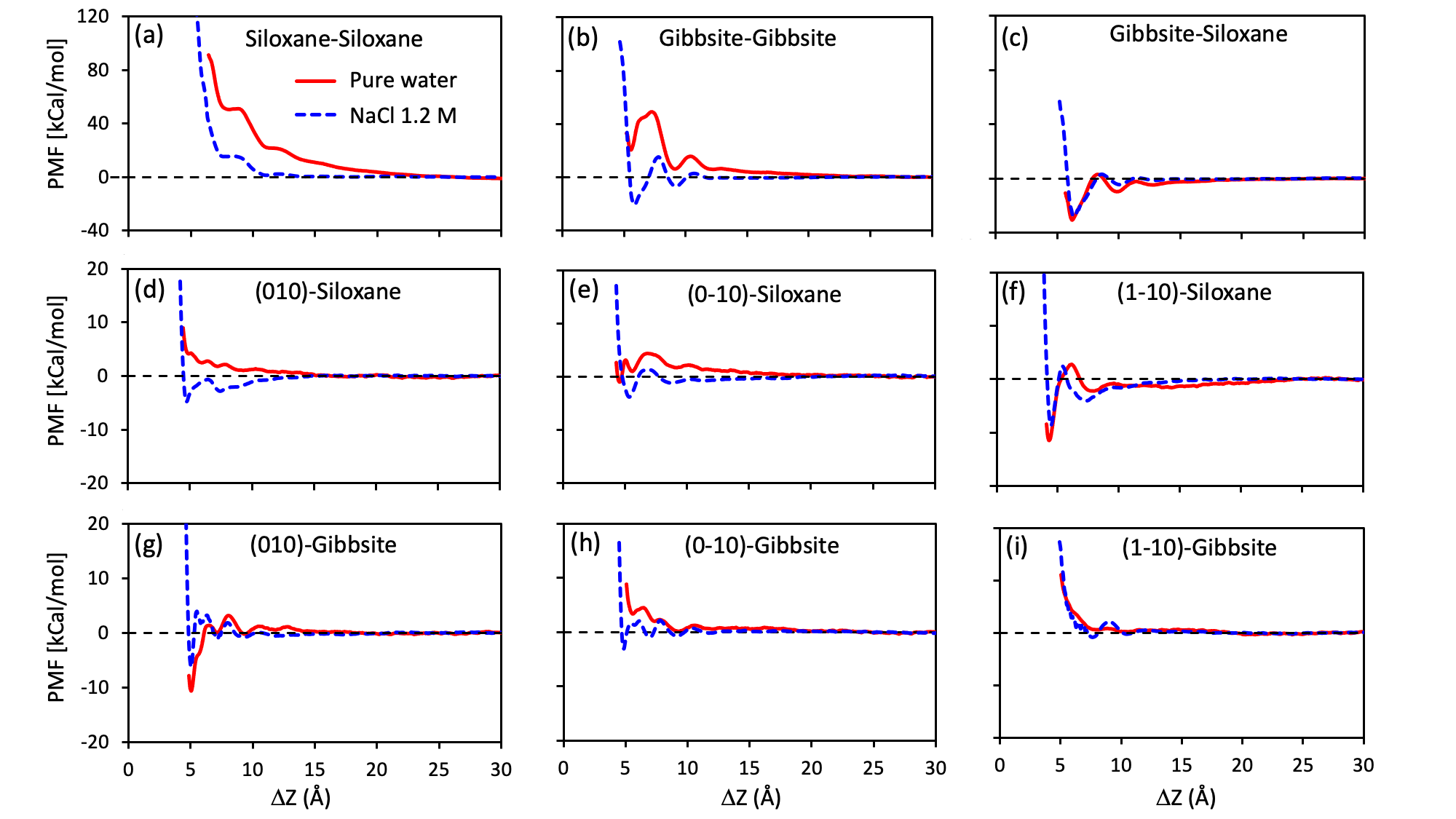}
\caption{\label{fig2}
Interaction between two kaolinite nanoparticles at different orientations.
Panels (a)-(c) are for face-to-face interfaces; panels (d)-(i) for edge-to-face interfaces.
Red and blue curves are the 
PMF profiles in pure water and in 1.2 M of sodium chloride solution, respectively. 
They are obtained from umbrella sampling simulations in the NVT ensemble at a temperature of 350~K, employing the ClayFF atomic force field for the kaolinite particles and the SPC/E water model. 
The abscissa, $\Delta$Z, is the minimum distance among the ``heavy'' atoms (i.e., Al or Si atomic species) of the two nanoparticles. 
}
\end{figure}

The PMF curves obtained from our simulations are shown in Figure \ref{fig2}.
We find that the PMF profiles are highly dependent on the relative orientation between the two particles. 
In particular, the gibbsite-siloxane orientation is the most attractive, whereas the gibbsite-gibbsite and siloxane-siloxane orientations are overall repulsive. This behaviour is explained qualitatively, at least for the pure water case, in terms of electrostatics, as a kaolinite nanoparticle forms a dipole with the siloxane surface negatively charged and the gibbsite surface positively charged. 
The face-to-edge orientations exhibited a highly variable pattern of behaviour, with two out of six orientations being weakly attractive and the remaining being overall repulsive. 
The face-to-edge PMFs are noticeably flatter than the face-to-face PMFs. This is only in part explained by the size of the interface: in our simulations the face-to-face orientation has an interface that is roughly 2.3 times larger than the face-to-edge, while the free energy barriers in the former are much larger than a factor 2.3; see the plot of the PMF per surface area reported in 
Figure~S6 
of the SI. 
It is likely that the atomically rough morphology of the edges prevents the formation of highly pronounced hydration layers, as observed in simulations conducted for hydrated crystalline vs amorphous silica, which could reduce the intensity of the resultant PMF profiles \cite{Argyris2009HydrationSubstrates, Fan2011AmphiphilicSimulations, Argyris2013HydrationSimulations}.

The presence of ions in solution shows a significant impact on PMF profiles. Ions screen the long-range electrostatic interaction between the two particles, especially in the face-to-face oriented interfaces. In particular, siloxane-siloxane and gibbsite-gibbsite orientations exhibit decreased repulsion at large distances when ions are in solution. 
The greatest effect is observed on the interactions between the particles along the gibbsite-gibbsite orientation, where the PMF profile is repulsive in pure water but becomes attractive in saline solution. In the case of gibbsite-siloxane, ions reduce slightly the attraction between the two particles. 
We observed the same trend for the effect of salt in the edge-to-face oriented systems, namely, that salt decreases long-range interactions between two particles. As a result, we found that in some systems the PMF changes from repulsive to attractive due to the presence of salts. 
It is worth noting that in our calculations for the face-to-face orientations, the PMF profiles are limited to a minimum distance of approximately 5 \AA. This is due to the fact that the free energy barrier is very high at closer distances \cite{Ho2021Molecular-levelWater}, preventing effective sampling of phase space \cite{Shen2021MolecularWater} even though the locations of the local minima and maxima can still be identified 
(see section~S6 
and Figure~S7 
of the SI).

\subsection{Short and long range characteristics of PMF profiles}\label{sec:interparticle}

The PMF profiles shown in Figure \ref{fig2} exhibit a common feature: they are very corrugated when the interface thickness $\Delta$Z is smaller than ca. 1.5~nm, and they are smoother for larger $\Delta$Z.
This allows us to distinguish between a short range regime, characterised by the presence of several local minima and free energy barriers among them in the PMF profile, and a relatively long regime.

\begin{figure}[ht!]
\centering
\includegraphics[width=\textwidth, trim=5.8cm 1cm 7cm 0cm,clip]{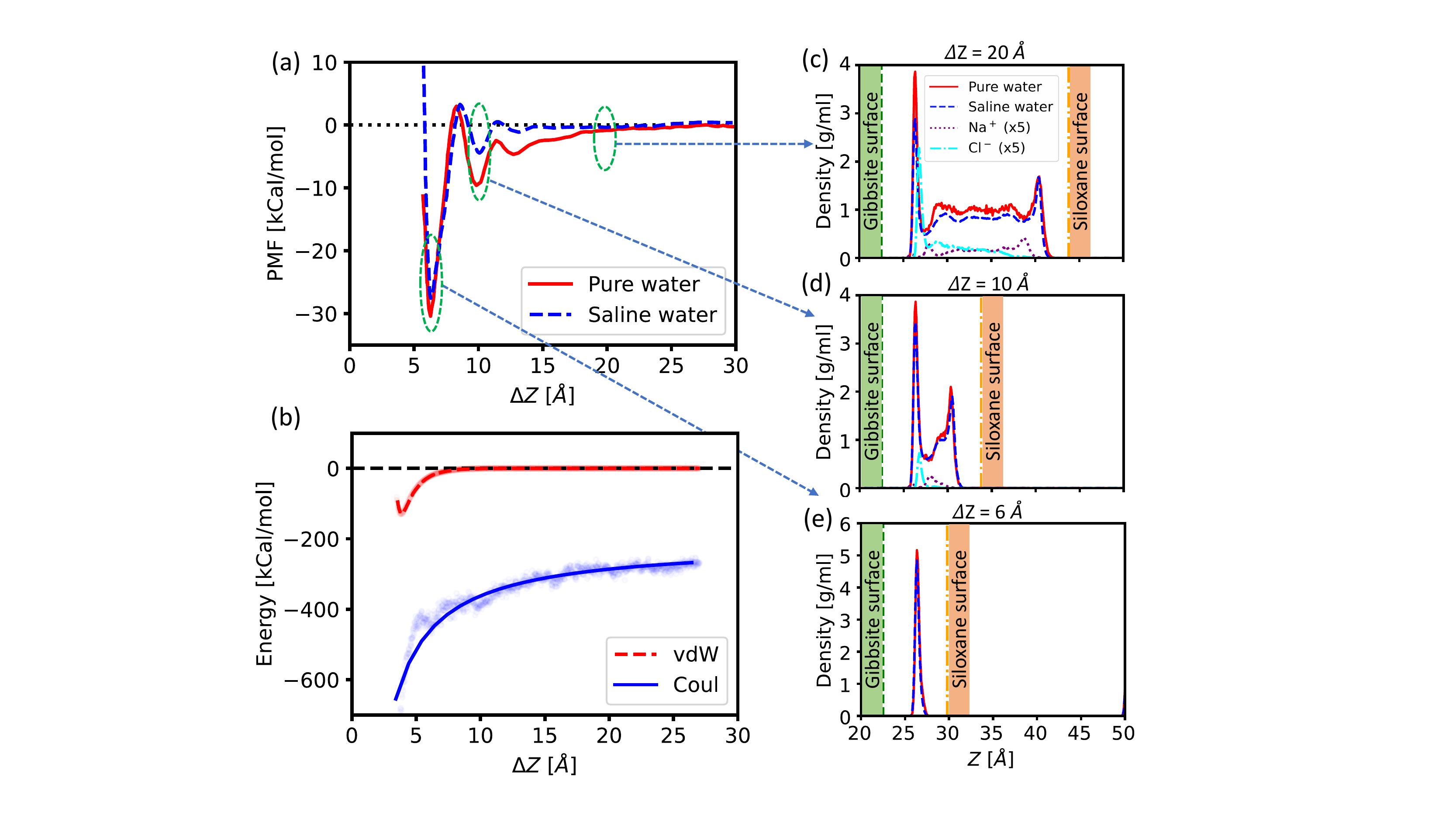}
\caption{\label{fig3}
Overview of the interaction and the hydration for interfaces of different thickness. 
(a) PMF profiles between two kaolinite particles in pure and saline water systems for the gibbsite-siloxane orientation. (b) Decomposition of particle-particle interactions: vdW and Coulomb potentials as a function of the distance between the two particles. (c), (d), and (e) Density profiles of water, Na$^+$, and Cl$^-$ along the Z direction of the simulation box calculated for the pure and saline water systems at the particle-particle distance of 20 \AA, 10 \AA, and 6 \AA, respectively.}
\end{figure}

The gibbsite-siloxane orientation shows the most intense effective attraction.
In this section we report a deeper analysis of this specific orientation.
Figure \ref{fig3}(a) zooms in on the gibbsite-siloxane PMF profiles obtained with pure and saline water. 
In both cases we clearly identify three local minima, the first in the case of pure water being at $\Delta$Z=6.2~\AA{}, the second at 9.9~\AA{}, and the third at 12.8~\AA{} (as we can precisely infer from the analysis of the US simulations, 
see Figure~S7 
of the SI). 
In almost the same position we have the minima for saline water \footnote{Notice that this almost exact match of the minima locations for the pure and saline water solution is observed in most the PMF profiles studied there.}.
The main difference between the PMF profiles obtained in the pure and saline water is that the latter appears having no long range interaction, and the intensity of the PMF minima and maxima change slightly due to the presence of salt.

The smoothness of the long range and the corrugation of the short range can be explained in terms of the structure of the water at the interface. 
%
When the interface is thick (say, the separation between the particles is larger than ca. 1.5~nm) the water in the middle of the interface has the density of bulk liquid water at the conditions considered.
See for instance the case of $\Delta$Z=2~nm in Figure \ref{fig3}(c).
In contrast, at distances shorter than 1.5~nm the simulations show that water at the interface forms hydration layers, in a number ranging from 1 to 3. See for instance the case of $\Delta$Z=1~nm in Figure \ref{fig3}(d), having a bilayer of water between the particles or a single layer in the case of $\Delta$Z=0.6~nm in Figure \ref{fig3}(e).
It shall be noticed that the presence of two or three water layers at the interface between liquid water and a solid surface is an expected feature on atomically smooth solid surfaces, observed also in experiments \cite{Bjorneholm2016, Perkin2004_confinedwater}.
In our system water is in fact confined in the region between the two nanoparticles, thus forming hydration layers on both the surfaces.
The minima in the PMF profiles correspond to the inter-particle distances that optimally accommodate 1, 2 or 3 layers, and the barriers between the minima arise because the corresponding particle-particle distances are energetically unfavourable.

The plots in Figure \ref{fig3}(c,d,e) show that the density of water at the interface is almost the same for pure and saline water, the latter being slightly lower due to the simultaneous presence of ions at the interface. However, at the simulated salinity the concentration of ions is almost negligible compared to that of water, so in the plots, the density of Na$^+$ and Cl$^-$ is magnified by a factor of 5.
We show in Figure \ref{fig3}(c,d) the different behaviour of ions at the two kaolinite faces: the gibbsite face shows a pronounced peak of Cl$^-$ anions, and the siloxane face shows a quite smaller peak of the Na$^+$ cations and the absence of any Cl$^-$ anion in its proximity.
Thus, ions distribute on the two kaolinite faces to counteract the intrinsic surface charge, screening the inter-particle Coulomb interaction, in agreement with literature \cite{Kumar2017, Papavasileiou2018, Vasconcelos2007}.
This screening effect appears as the reason for the lack of long range interaction among the kaolinite particles in saline solution observed in Figure \ref{fig3}(a).

To illustrate that the long range attraction of the two particles is indeed due to the electrostatic interaction, we have disentangled some of the contributions that sum up to determine the overall interaction energy, and specifically the van der Waals (vdW) and the Coulomb inter-particle contributions \footnote{ In this contribution we are just considering the two particles and ignoring anything else (water and ions). }.
The mean vdW and Coulomb contributions are plotted in Figure \ref{fig3}(b) as a function of the distance $\Delta$Z.
The van der Waals potential exhibits a short-range effect, becoming negligible beyond 1~nm. 
The electrostatic potential exhibits long-range features, as expected. Therefore, at distances exceeding 1~nm 
the interparticle interaction is mostly due to electrostatics. We observed distinct differences in the interactions depending on the orientation of the particles. When the two approaching surfaces are dissimilar (i.e., gibbsite-siloxane), both the van der Waals and electrostatic potentials yield attractive interactions. In contrast, when the approaching surfaces are identical, e.g., gibbsite-gibbsite and siloxane-siloxane, we observe a repulsive behaviour (see 
Figure~S8 
of the SI). This is due to the identical surface charges on the two approaching particles. 
As discussed above, in saline water the ions deposit near the surfaces screening the long range Coulomb attraction/repulsion. 

\subsection{The role of the hydration film}\label{sec:hydration}

Based on the behaviour of the inter-particle interaction energy, see in Figure \ref{fig3}(b),
the interparticle interactions should yield smooth PMF curves. This is not the case, as observed in the previous section, and the 
PMF profiles exhibit a corrugated shape at distances $\Delta$Z smaller than ca. 1.5nm, see Figure \ref{fig2}. The solvent plays an important role in determining this corrugation
\cite{Argyris2013HydrationSimulations, Lin2019Ion-SpecificInterfaces, Ho2021Molecular-levelWater}. 
In our study, we found that when two particles approach one another, their hydration films tend to migrate closer together and merge into a single layer at short distances. 
We believe that the compatibility of the two hydration films correlates with the energetic and enthalpic cost of the particle-particle aggregation. 

\begin{figure}[ht!]
\centering
\includegraphics[width=\textwidth]{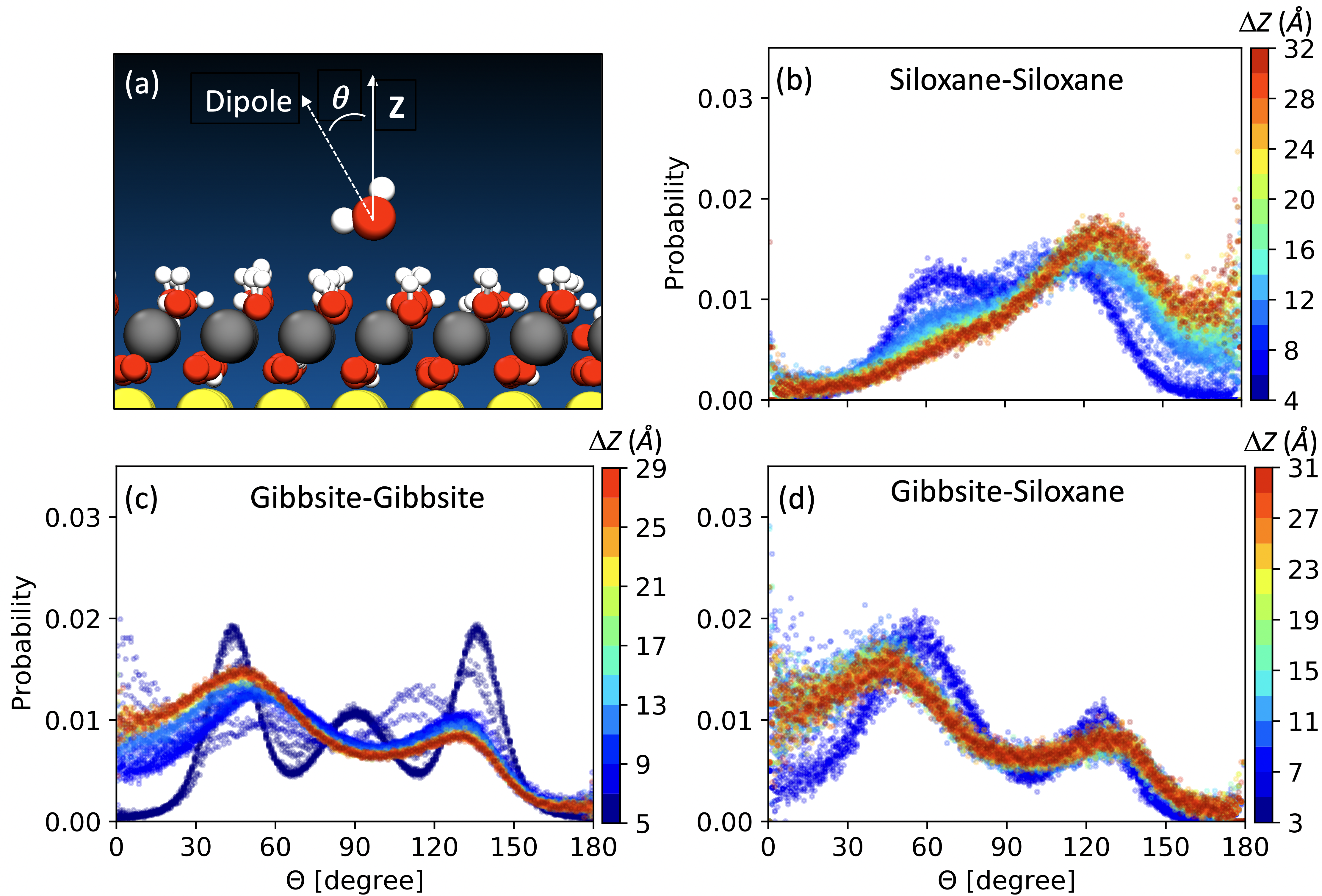}
\caption{\label{fig4}
Orientation of water molecules at the interface. 
(a) Definition of the angle formed between water dipole moments and the Z direction of the simulation box. (b)-(d) Distribution of the angle formed between the dipole vector of water molecules in the first hydration layer and the Z direction of the simulation box. Colour scheme represents the distributions at various particle-particle separations.}
\end{figure}

More specifically, the aggregation of two particles at a specific orientation should be favourable if the structure of the water confined between the particles is similar to the structure of interfacial water when the particles are far away, and should be unfavourable otherwise.
Thus, we evaluated the distribution of the orientation of the water molecules in the first hydration layer of the bottom particle in out MD simulations. 
A comparison of the distributions obtained for the different orientations and at different distances $\Delta$Z are reported in Figure \ref{fig4}.
Panels (b, c, d) 
show the results of the water orientation distribution. As the two particles approach one another in the siloxane-siloxane and gibbsite-gibbsite orientations, the orientation of water molecules in the first hydration layer undergoes a significant change. In contrast, in the case of gibbsite-siloxane orientation, the distribution at various distances changes only slightly as the two particles approach each other. 
Similar observations are drown if we consider the  1-dimensional density profiles for water hydrogen and oxygen atoms along the Z direction of the simulation box for the three orientations, 
see Figure~S10
(a)-(i) and section~S9 
of the SI. 
Thus, for the gibbsite-siloxane orientation, the energy cost associated with changing the water orientation is expected to be small, different from siloxane-siloxane and gibbsite-gibbsite orientations. 
The above conclusion is supported also by the behaviour of the solvation free energy as a function of the interparticle separation, which we have evaluated from our MD simulations and shown in 
Figure~S9 
of the SI (details are in 
section~S8 
of the SI).

\begin{figure}[ht!]
\centering
\includegraphics[page=1, width=\textwidth, trim=0cm 0cm 0cm 0cm,clip]{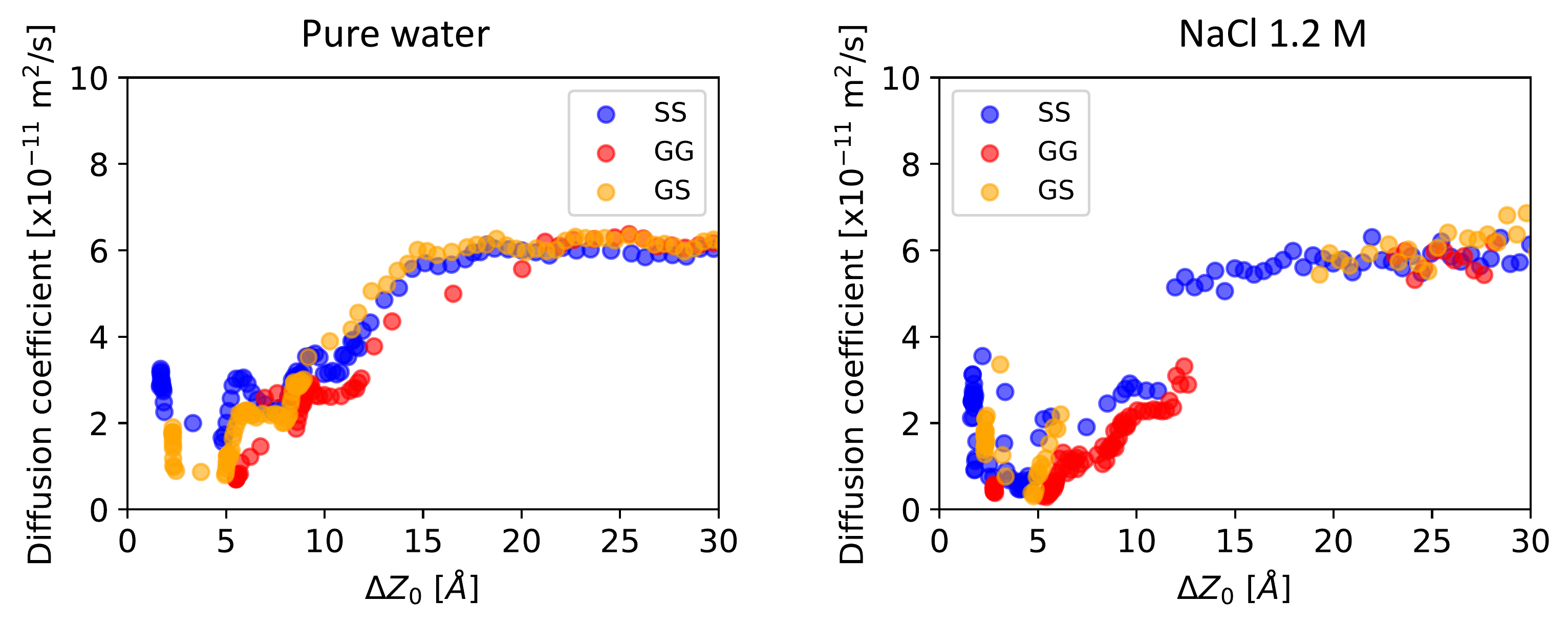}
\caption{\label{fig_particle_diffusion}
Diffusion coefficients along the Z direction of the upper kaolinite particle in the simulation box as a function of particle-particle separation calculated for different orientations in pure water (left) and in saline water at 1.2 M NaCl concentration (right).
}
\end{figure}

The hydration film has also an important effect on diffusion properties of the approaching nanoparticles. 
It is well know that nanoparticles in solution will exhibit a Browninan motion, with a diffusion coefficient proportional to the temperature of the system and to the inverse of the nanoparticle size, as prescribed by the Stokes-Einstein relation.
Is the diffusion coefficient of a nanoparticle affected by the presence of a second nanoparticle? In which way?
We evaluated the diffusion coefficient of one nanoparticle at a given distance from the other nanoparticle in our simulations using the approach of Ref. \cite{Woolf1994ConformationalEffects, Hummer2005Position-dependentSimulations, Phan2016MolecularPerformance} and described in 
section~S13 
of the SI.
Figure \ref{fig_particle_diffusion} shows the diffusion coefficients for each of the three face-to-face orientations, both for pure and saline water.
At an interparticle distance $\Delta$Z larger than 1.5~nm the diffusion coefficient reaches a plateau of ca. $6\times 10^{-11} m^2/s$, which appears as the bulk value of the diffusion coefficient.
The bulk value appears the same on pure and saline water.
Closer than 1.5~nm the diffusion coefficient decreases, becoming as slow as more than six times smaller than in the bulk.
The trend of the diffusion coefficient as a function of the distance  $\Delta$Z appears only weakly affected by the specificity of the face-to-face orientation and on the presence or absence of salt in water.
However, it can be noticed that in saline water the decrease of mobility is more pronounced that in pure water.

This reduction of the nanoparticle mobility can be explained from the behaviour of the hydration water. We have measured in our simulations a sharp reduction of the diffusion coefficient of water in the proximity of the nanoparticle faces, see 
section~S12 
and Figure~S13 
of the SI.
Notice that this decrease of mobility of water, and the corresponding increase in viscosity, is in good agreement with several experimental observations \cite{Raviv2001FluidityFilms, Perkin2004_confinedwater}.
Moreover, experiments \cite{Phillips1980ViscosityPressures, Kestin1981TablesMPa} also show that the viscosity of water increases by roughly 10 percent with the concentration of NaCl used in our study.
The reduction of the nanoparticle mobility when closer than 1.5 nm appears to be closely related with the increased viscosity of hydration film. 

\section{Conclusions}\label{sec:conclusions}

In this work we have used atomistic molecular dynamics simulations and enhanced sampling techniques to investigate and characterise the interaction between two nanoparticles of kaolinite.
Kaolinite particles have roughly the shape of an hexagonal prism, with the two basal faces being inequivalent, as are the side edges.
As such, the interaction between two particles depends on the relative orientation. We studied the three possible face-to-face orientations and all the representative face-to-edge orientations.

A remarkable anisotropy of the interaction emerged.
Face-to-edge interfaces yield a potential of mean force that is much flatter than the face-to-face interfaces. In the latter case two of the orientations result in a repulsive interaction and one, the interface between the inequivalent faces, yields an attractive interaction.
We notice that the PMF profiles exhibit different features depending on the distance between the two particle surfaces.
On the one hand, by disentangling the contributions that sum up to give the overall interaction energy, we infer that the attractive or repulsive nature of the interaction, for distances larger than ca. 1.5~nm, is determined mainly by electrostatics. 
On the other hand, at shorter separations the structure of water confined between the particles has to be carefully understood, as this determines the interaction profile. We notice a tendency to form water layers, which correspond to local minima in the free energy which yields a very corrugated PMF.

We have evaluated the PMF profiles in pure water and in saline solution. In saline solution the ions have a tendency to adsorb preferentially on the basal faces of the particles, counteracting the intrinsic dipole across  the kaolinite particles. 
Therefore, ions screen the electrostatic inter-particle interaction, and the long range PMF profiles show a much weaker interaction in saline water than in pure water. 
However, ions do not appear to affect greatly the corrugation of the PMF profiles at small particle-particle separations.

Our characterisation of the interaction between kaolinite nanoparticles can be compared with alternative theoretical approaches to characterise interaction energies between (nano)-particles.
The most common approach is the DLVO theory, which is in general valid at large particle-particle separations. 
We have here investigated the PMF profile up to 3~nm, so our approach can be complementary to DLVO, as it assesses the properties of interfaces among particles at larger distances.
The interval of the PMF profile that we have investigated here is, we think, the most relevant for the aggregation of particles and the formation of fines. It can be used as a solid basis for further investigations aimed at modelling the aggregation of several particles. 

Experiments based on high-resolution atomic force microscopy, have already provided important information on the properties of clay interfaces \cite{Gupta2010, Gupta2011, Liu2014, Siretanu2014, Kumar2016, Kumar2017}. They are mostly interpreted in the content of DLVO theory and used to estimate the surface charge distribution in clays at different enviromental conditions.
Hopefully, new experiments performed between isolated kaolinite particles and functionalised tips (including surface charge measurements) could validate the results presented here and provide further insights into the importance of the ionic environment on particle aggregation.

\section*{acknowledgements}
\noindent 
We thank Stephen J. Cox and Andrey G. Kalinichev for insightful discussions. 
AZ acknowledges financial support from the Leverhulme Trust, grant number RPG-2020-038.
This work was also supported by BP Exploration Operating Company Limited, University College London/The Thomas Young Centre, and Innovative UK under the Knowledge Transfer Partnership, number KTP011009.
We are also grateful, for computational resources, 
to the BP HPC facilities,
to ARCHER UK National Supercomputing Service, the United Kingdom Car Parrinello (UKCP) consortium (EP/ F036884/1), 
the London Centre for Nanotechnology and 
University College London Research Computing,
the UCL Myriad and Kathleen High Performance Computing Facility (Myriad@UCL, Kathleen@UCL), 
the U.K. Materials and Molecular Modelling Hub for computational resources, which is partially funded by EPSRC (EP/P020194/1 and EP/ T022213/1), 
the CPU hours by CSCS under Project ID s1000, 
and the Cambridge Service for Data Driven Discovery (CSD3) operated by the University of Cambridge Research Computing Service (www.csd3.cam.ac.uk), provided by Dell EMC and Intel using Tier-2 funding from the Engineering and Physical Sciences Research Council (capital grant EP/P020259/1), and DiRAC funding from the Science and Technology Facilities Council (www.dirac.ac.uk).

\bibliography{ref}

\end{document}